\documentclass[letter]{aa}
\usepackage{epsfig}
\usepackage{natbib}
\bibpunct{(}{)}{;}{a}{}{,}

\newbox\grsign \setbox\grsign=\hbox{$>$} \newdimen\grdimen \grdimen=\ht\grsign
\newbox\simlessbox \newbox\simgreatbox \newbox\simpropbox \newbox\wtildebox 
\setbox\simgreatbox=\hbox{\raise.5ex\hbox{$>$}\llap
     {\lower.5ex\hbox{$\sim$}}}\ht1=\grdimen\dp1=0pt
\setbox\simlessbox=\hbox{\raise.5ex\hbox{$<$}\llap
     {\lower.5ex\hbox{$\sim$}}}\ht2=\grdimen\dp2=0pt

\newcommand{\be}{\mbox{\begin{equation}}}
\newcommand{\ee}{\mbox{\end{equation}}}

\newcommand{\tdis}{\mbox{$t_{\rm dis}^{\rm total}$}}

\newcommand{\Cref}{\mbox{$m_{\rm ref}$}}

\newcommand{\msun}{\mbox{M$_\odot$}}

\renewcommand{\d}{{\rm d}} 
\newcommand{\mc}{\mbox{$M_{\rm cl}$}}
\newcommand{\mci}{\mbox{$M_{\rm cl,i}$}}
\newcommand{\mct}{\mbox{$M_{\rm cl}(t)$}}

\newcommand{\ml}{\mbox{$M_{\odot}~L_{\odot}^{-1}$}}

\begin{document}

\title{Explaining the mass-to-light ratios of globular clusters
}

\author{J.~M.~Diederik~Kruijssen \inst{1,2}}

\institute { 
                 {\inst{1}Astronomical Institute, Utrecht University, 
                 PO Box 80000, NL-3584CC Utrecht, The Netherlands\\
                 (e-mail: {\tt kruijssen@astro.uu.nl})}\\
                 {\inst{2}Leiden Observatory, Leiden University,
                 PO Box 9513, NL-2300RA Leiden, The Netherlands}
            }

\date{Received 22 May 2008 / Accepted 4 June 2008}

\offprints{J.~M.~Diederik~Kruijssen, e-mail: {\tt  kruijssen@astro.uu.nl}}

\abstract{The majority of observed mass-to-light ratios of globular clusters are too low to be explained by `canonical' cluster models, in which dynamical effects are not accounted for. Moreover, these models do not reproduce a recently reported trend of increasing $M/L$ with cluster mass, but instead predict mass-to-light ratios that are independent of cluster mass for a fixed age and metallicity.}
{This study aims to explain the $M/L$ of globular clusters in four galaxies by including stellar evolution, stellar remnants, and the preferential loss of low-mass stars due to {{energy equipartition}}.}
{Analytical cluster models are applied that account for stellar evolution and dynamical cluster dissolution to samples of globular clusters in Cen A, the Milky Way, M31 and the LMC. The models include stellar remnants and cover metallicities in the range $Z=0.0004$---$0.05$.}
{Both the low observed mass-to-light ratios and the trend of increasing $M/L$ with cluster mass can be reproduced by including the preferential loss of low-mass stars, {{assuming}} reasonable values for the dissolution timescale. This leads to a mass-dependent $M/L$ evolution and increases the explained percentage of the {{observations}} from 39\% to 92\%.}
{This study shows that the hitherto unexplained discrepancy between observations and models of the mass-to-light ratios of globular clusters can be explained by dynamical effects{{, provided that the globular clusters exhibiting low $M/L$ have dissolution timescales within the ranges assumed in this Letter}}. Furthermore, it substantiates that $M/L$ cannot be assumed to be constant with mass at fixed age and metallicity.}


\authorrunning{J.~M.~Diederik~Kruijssen}
\titlerunning{Explaining the mass-to-light ratios of globular clusters}

\maketitle


\section{Introduction} \label{sec:intro}
The mass-to-light ratios of globular clusters (GCs) have been given a lot of attention recently \citep[e.g.,][]{mclaughlin05,rejkuba07,mieske08a,dabringhausen08}. \citet{rejkuba07} have observed an $M/L$ trend with cluster mass above a certain cluster mass, as more massive clusters appear to have higher $M/L$ than low-mass clusters \citep[see also][]{mandushev91}. This is an observation contrary to fundamental plane studies of GCs \citep[e.g.,][]{mclaughlin00} and also in strong disagreement with the constant $M/L$ for fixed age that is commonly assumed in observational and theoretical GC studies \citep[e.g.,][]{harris06,mora07,bekki07}. Moreover, for Galactic GCs \citet{mclaughlin00} find $M/L_V=1.45~\ml$, whereas Simple Stellar Population models \citep[e.g.,][]{bruzual03,andersfritze03} predict significantly higher values of $M/L_V=2$---$4~\ml$ for typical GC metallicities. Given the important role of GCs in galactic astronomy, it is essential to explain these apparent contradictions.

In numerical and analytical studies of dynamical effects in clusters \citep[e.g.,][]{baumgardt03,lamers06,kruijssen08} it has become clear that the dynamical evolution of clusters strongly affects cluster luminosity, colour and mass-to-light ratio. In \citet[hereafter KL08]{kruijssen08} it is shown how the evolution of these observables changes due to dynamical effects such as the preferential loss of low-mass stars and the retain of stellar remnants, but also due to the stellar initial mass function and metallicity. It is shown that $M/L$ cannot be assumed to be constant for a fixed age and metallicity, but {instead} depends on cluster mass when dynamical effects are accounted for.

In this Letter, the analytical cluster models from KL08 are applied to explain the observations of GCs in several galaxies from \citet{rejkuba07} and \citet{mieske08}. In Sect.~\ref{sec:model} I first summarise the models presented in KL08, which is applied to the observations in Sect.~\ref{sec:rejkuba}. In Sect.~\ref{sec:mlplane} the effect of metallicity and the cluster dissolution timescale on cluster evolution in the \{$M,M/L_V$\}-plane is investigated. The observations are discussed in Sect.~\ref{sec:obs} and are compared to the models in Sect.~\ref{sec:expl}. A discussion of the results and the conclusions are presented in Sect.~\ref{sec:concl}.

\section{Modeling method} \label{sec:model}
In this study, analytic cluster models are used that incorporate the effects of stellar evolution, stellar remnant production, cluster dissolution and {energy equipartition}. They are summarised here and are treated in more detail in KL08.

In the models, clusters gradually lose mass due to stellar evolution and dissolution. The total cluster mass evolution is described by $\d\mc/\d t=(\d\mc/\d t)_{\rm ev}+(\d\mc/\d t)_{\rm dis}$, with the first term denoting stellar evolution and the second representing dissolution. Taking into account the formation of stellar remnants and the mass-dependent loss of stars by dissolution, this provides a description of the changing mass function and cluster mass in remnants.

Stellar evolution is included by using the Padova 1999 isochrones\footnote{These isochrones are based on \citet{bertelli94}, but use the AGB treatment as in \citet{girardi00}.}. It removes the most massive stars from the cluster and increases the dark cluster mass by turning stars into remnants, which is included by assuming an initial-remnant mass relation. A \citet{kroupa01} IMF is assumed.

Cluster dissolution represents the dynamical cluster mass loss due to stars passing the tidal radius. This mass loss acts on a dissolution timescale $\tau_{\rm dis}$:
\begin{equation}
\label{eq:dmdtdis}
  \left(\frac{\d \mct}{\d t}\right)_{\rm dis} = -\frac{\mct}{\tau_{\rm dis}} = -\frac{\mct^{1-\gamma}}{t_0} ,
\end{equation}
where $\tau_{\rm dis}$ is related to the present day cluster mass $\mct$ as $\tau_{\rm dis}=t_0\mct^\gamma$ \citep{lamers05}, leading to the second equality in Eq.~\ref{eq:dmdtdis}. The characteristic timescale $t_0$ depends on the environment and determines the strength of dissolution. For example, in the case of tidal dissolution $t_0$ depends on tidal field strength and thus on galactocentric radius and galaxy mass. Typical values are $t_0=10^5$---$10^8$~yr \citep[e.g.,][]{lamers05a}, translating into a total disruption time $\tdis\approx 10^8$---$10^{11}$~yr for a 10$^5~\msun$ cluster. From $N$-body simulations of tidal dissolution \citep{lamers05a} and observations \citep{boutloukos03,gieles05a}, the exponent $\gamma$ is found to be $\gamma\approx 0.62$.

The effect of dissolution on the mass function {depends on} the dynamical state of the cluster. If it {has reached energy equipartition}, i.e., after core collapse, {the cluster becomes mass-segregated and low-mass stars are preferentially lost}. This occurs at about 20\% of the total cluster lifetime \citep{baumgardt03}. For clusters without {equipartition}, bodies of all masses are lost with similar probabilities.

Cluster photometry is computed by integrating the stellar mass function over the {stellar} isochrones, yielding cluster magnitude evolution $M_{\rm \lambda}(t,\mci)$ for a passband $\lambda$ and a cluster with initial mass $\mci$.

\section{Applying the cluster models to observed clusters} \label{sec:rejkuba}
In this section I present the evolution of clusters in the \{$M,M/L_V$\}-plane, and apply this to explain the \{$M,M/L_V$\} distribution observed in real clusters.

\subsection{Cluster evolution in the mass-$M/L_V$ plane} \label{sec:mlplane}
In `canonical' Simple Stellar Population (SSP) models, clusters only evolve due to stellar evolution, and therefore their mass-to-light ratios do not change due to dynamical effects. As the most massive stars (with low $M/L_V$) gradually disappear, in these models $M/L_V$ is a monotonously increasing function of time that is constant for any set of clusters at a single age and metallicity. However, this is only correct if cluster dissolution occurs independently of stellar mass and the shape of the stellar mass function is preserved, i.e., {there is {\it no} preferential loss of low-mass stars}.

Contrary to clusters from SSP models, in reality clusters do {preferentially lose low-mass stars} \citep[e.g.,][]{hillenbrand98,albrow02,baumgardt03}. KL08 have shown that this strongly affects the $M/L$ evolution of clusters due to the preferential loss of low-mass stars (having high $M/L$), and that consequently $M/L$ cannot be assumed to be constant for clusters of a given age.

\begin{figure*}[t]\centering
\resizebox{16cm}{!}{\includegraphics{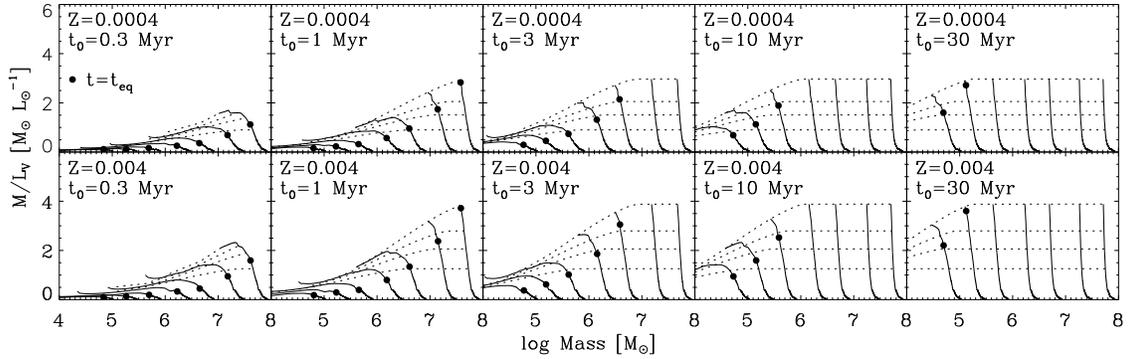}}
\caption[]{\label{fig:model}
   Cluster evolution in the \{$M,M/L_V$\}-plane for metallicities $Z=\{0.0004,0.004\}$ ($[{\rm Fe/H}]=\{-1.7,-0.7\}$) and dissolution timescales $t_0=\{0.3, 1,3,10, 30\}$~Myr. Solid lines represent evolutionary tracks for initial masses in the range $\mci=10^5$---$10^8$~\msun~with 0.5-dex intervals, while the dotted curves denote cluster isochrones at ages $t=\{4,8,12,19\}$~Gyr. {Dots denote the onset of {the preferential loss of low-mass stars} for each evolutionary track.}
    }
\end{figure*}
In order to explain the appearance of clusters in the \{$M,M/L_V$\}-plane, I first compute the model cluster evolutionary tracks for different metallicities and dissolution timescales. The results are shown in Fig.~\ref{fig:model}.

Initially, {model} clusters are not {in equipartition} and they evolve to lower masses and {\it increasing} $M/L_V$ (due to the death of massive stars), moving up and to the left in Fig.~\ref{fig:model}. From the moment they reach mass segregation, happening earlier for low-mass clusters {due to quicker relaxation} \citep{baumgardt03}, they preferentially lose low-mass stars, which have high $M/L$. This {\it decreases} the cluster $M/L$ and explains the maximum in the cluster evolution curves for lower initial masses. Along the cluster isochrones of constant age (dotted lines in Fig.~\ref{fig:model}), $M/L_V$ increases with mass, since at any given age low-mass clusters have spent more time in {energy equipartition} and thus have retained more massive (i.e., low-$M/L$) stars compared to massive clusters. The {cluster} isochrones flatten at the highest masses, since these clusters have yet to reach {equipartition}, leaving them at constant $M/L$.

Within a galaxy, its GCs generally have {similar} ages \citep[e.g.,][]{vandenberg90}. Observations of GC systems should thus approximately follow {cluster} isochrones in the \{$M,M/L_V$\}-plane. Therefore, the isochrones are used in Fig.~\ref{fig:model} to probe the influence of metallicity and dissolution timescale on the expected GC distribution. Increasing metallicity leads to a higher maximum $M/L$, and thus also to steeper isochrones. Increasing the dissolution timescale shifts the entire isochrone to the left: for long dissolution timescales, it takes more time to reach {equipartition} and therefore only clusters of the lowest masses {have preferentially lost low-mass stars}. The dissolution timescale thus sets the location of the `knee' in the isochrones, which is the cluster mass at which they flatten due to the absence of {equipartition}. This can be used to determine the dissolution timescale range of an observed GC system.

\subsection{Observations of globular cluster mass-to-light ratios} \label{sec:obs}
\begin{figure}[t]\centering
\resizebox{8.5cm}{!}{\includegraphics{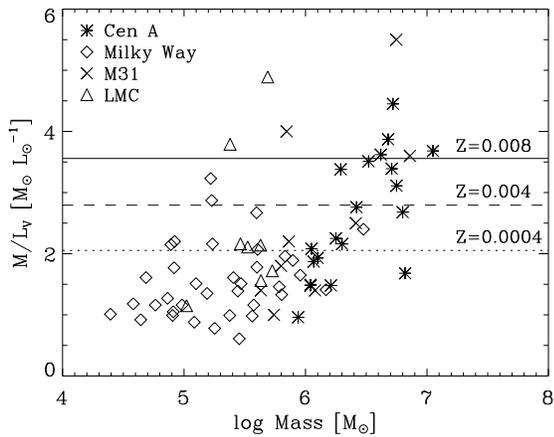}}
\caption[]{\label{fig:obs}
Mass-to-light ratio versus mass for globular clusters in different galaxies. The Cen A data is from \citet{mieske08}, all other data is taken from \citet{rejkuba07}. The canonical (mass-independent) models are overplotted as solid, dashed and dotted lines for $Z=\{0.0004,0.004,0.008\}$, respectively.
    }
\end{figure}
In \citet[Fig.~9]{rejkuba07} the {(dynamical-)} \{$M,M/L_V$\}-plane is presented for GCs in several host galaxies. In Fig.~\ref{fig:obs}, I show their \{$M,M/L_V$\}-plane for GCs from Cen~A, the Milky Way, M31, and the Large Magellanic Cloud (LMC). For the Cen~A data, aperture corrections as in \citet{hilker07} have been computed for the cluster masses and $M/L_V$ \citep{mieske08}. The $M/L_V$ values for {mass-independent} cluster models {(the `canonical' models from KL08)} are overplotted, and fail to reproduce a large part of the data. When considering the metallicities and errors of the data (not shown in Fig.~\ref{fig:obs}), only 39\% of the observed GCs can be covered within their 1-$\sigma$ errors if the canonical models are used\footnote{SSP models \citep[e.g.,][]{bruzual03,andersfritze03} all predict $M/L=2$---$4~\ml$ for GC metallicities, a Kroupa IMF and $t=12$~Gyr, comparable to the `canonical' models used here.}.

Although the data show quite some scatter, there {are indications for} a trend of increasing $M/L_V$ with mass{, such as the lack of low-$M/L$ clusters for $\log{M}>6.3$}. I argue that this is the same behaviour as can be observed in the {cluster} isochrones {including the preferential loss of low-mass stars} in Fig.~\ref{fig:model}, implying that the increase of $M/L_V$ with mass corresponds to a decreasing effect of {energy equipartition}.

\subsection{Explaining the mass-to-light ratios of globular clusters} \label{sec:expl}
\begin{figure*}[t]\centering
\resizebox{12cm}{!}{\includegraphics{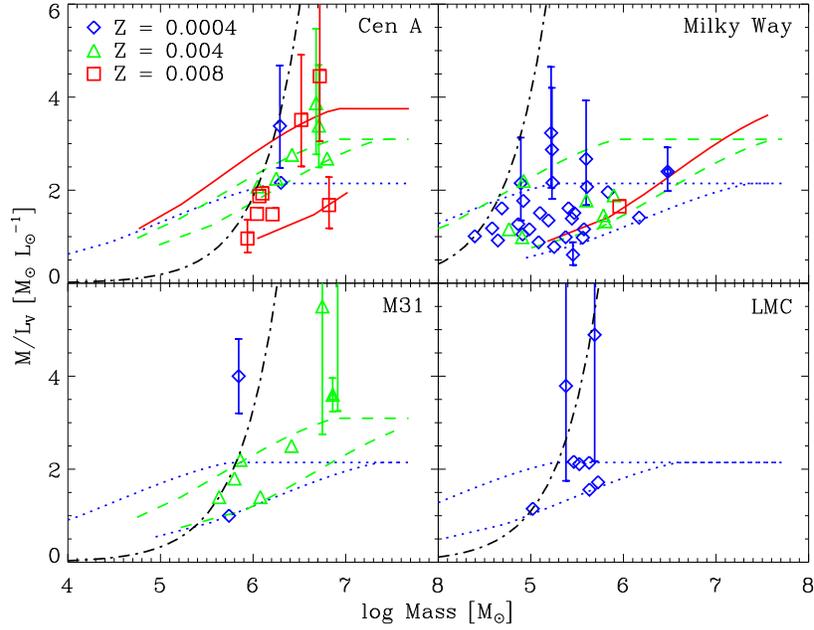}}
\caption[]{\label{fig:expl}
   Comparison of $t=12$~Gyr model cluster isochrones in the \{$M,M/L_V$\}-plane to globular clusters in four different galaxies. Clusters with metallicities $Z=\{0.0004,0.004,0.008\}$ are denoted with blue diamonds, green triangles, and red squares, respectively. Model curves for these metallicities are shown in the same colours, with dotted, dashed and solid lines, respectively. In all cases an age of $t=12$~Gyr is assumed. For each galaxy, a dash-dotted line of constant luminosity represents the faintest cluster that is covered by the models, illustrating that the samples are magnitude-limited. Metallicities are from \citet[Cen A]{beasley08}, \citet[Milky Way]{harris96}, \citet[M31]{dubath97} and \citet[LMC]{mackey04}. To prevent crowding, error bars are only shown for clusters that fall outside the range covered by the models.
    }
\end{figure*}

Figure~\ref{fig:expl} {compares} the model {cluster} isochrones ($t=12$~Gyr) to the GC data for Cen~A, the Milky Way, M31 and the LMC. The different colours represent three metallicities, and coloured model lines belong to data points of the same colour.
\begin{table}[tb]\centering
\begin{tabular}{|c c c c|}
  \hline \hline & & $t_0$ range (Myr) & \\ \hline \hline
  Galaxy & $Z=0.0004$ & $Z=0.004$ & $Z=0.008$ \\ \hline
  Cen A & $\leq 5$---5 & $\leq 1$---2 & {\bf 0.2}---2 \\
  M31 & $\leq 1$---10 & $\leq 0.5$---2 & \\
  MW & {\bf 1}---20 & {\bf 0.7}---8 & {\bf 0.6}---0.6 \\
  LMC & $\leq 3$---20 & & \\\hline
\end{tabular}
\caption[]{\label{tab:expl}
      {Required} dissolution timescale ranges for globular clusters of three metallicities in four studied galaxies, as derived from the cluster samples. Due to possible incompleteness {at low masses and high $M/L$ (see Fig.~\ref{fig:expl})}, all upper limits represent minimum values and lower limits often represent maxima. Limits that do not suffer from incompleteness are shown in boldface.
    }
\end{table}

For any galaxy and metallicity, the data cover an area in the \{$M,M/L_V$\}-plane that can be spanned by two model curves of different dissolution timescales. Left curves denote upper limits, while right curves represent lower limits for the dissolution timescale ranges in which clusters are observed. {These limits are chosen as such that they encompass the data points.} Contrary to the sparse coverage of the data by the canonical models (see Fig.~\ref{fig:obs}), it is shown in Fig.~\ref{fig:expl} that 92\% of the data can be explained within their 1-$\sigma$ errors by using the new models (KL08) that account for dynamical effects. The remaining 8\% has too high $M/L$ to be explained by stellar population models, unless their observed ages or metallicities are underestimated.

The minimum and maximum dissolution timescales that are required to explain the observations are summarised in Table~\ref{tab:expl}. All values fall within the physically reasonable range of 10$^5$---10$^8$~yr \citep{lamers05a}. For each galaxy, {a broad range of dissolution timescales is required to explain the observed range of $M/L$}. This is not surprising, since the observed clusters are located at various galactocentric radii and thus experience different tidal dissolution strengths. Regardless of this spread, there is a clear trend of decreasing {required} dissolution timescale with metallicity. This is likely to be related to the radial metallicity gradient observed in galaxies \citep[first established by][]{searle71}, with metal-rich clusters at small galactocentric radii and thus at short dissolution timescales. Another trend is that of decreasing {required} dissolution timescale with galaxy mass. Again, this is not surprising, as more massive galaxies {can} have stronger tidal fields and thus give rise to more rapid cluster dissolution.

\section{Discussion} \label{sec:concl}
In this Letter, I have shown that the hitherto unexplained discrepancy between observations and models of the mass-to-light ratios of globular clusters can be explained by dynamical effects. The preferential loss of low-mass stars {due to energy equipartition} gives rise to $M/L$ evolution that depends on cluster mass, contrary to what is assumed in canonical cluster models. This is confirmed by the application of models that include dynamical effects to the GC populations of Cen A, the Milky Way, M31 and the LMC.

Without {the preferential loss of low-mass stars}, current stellar population models cannot explain mass-to-light ratios below 2~\msun~L$^{-1}_\odot$ for metallicity $Z=0.0004$ and below 2.8~\msun~L$^{-1}_\odot$ for $Z=0.004$. As becomes clear from Fig.~\ref{fig:expl}, this would leave half of the cluster sample in Cen A and most of the Milky Way sample unexplained. Accounting for the effects of {energy equipartition} increases the explained percentage of the observations from 39\% to 92\%.

The dissolution timescales required to explain the observed GC samples lie within the physically reasonable range of $t_0=10^5$---10$^8$~yr. Observed trends of decreasing dissolution timescale with galaxy mass and metallicity are as expected when considering the strength of tidal dissolution and the radial metallicity gradient in galaxies.

The dependence of $M/L$ on cluster mass (and thus on luminosity) implies that photometrically derived masses using canonical models may be strongly overestimated (KL08). The results presented here underline the importance of accounting for dynamical effects when modeling clusters or interpreting observations of (globular) clusters.

\begin{acknowledgements}
I am very grateful to Marina Rejkuba for kindly providing me with an electronic version of the data presented in \citet{rejkuba07} and \citet{mieske08}. I also thank Steffen Mieske, Mark Gieles and the anonymous referee for constructive comments and Henny Lamers for encouragement, advice and suggestions.
\end{acknowledgements}

\bibliographystyle{aa}
\bibliography{mybib}

\begin{thebibliography}{31}
\expandafter\ifx\csname natexlab\endcsname\relax\def\natexlab#1{#1}\fi

\bibitem[{{Albrow} {et~al.}(2002){Albrow}, {De Marchi}, \& {Sahu}}]{albrow02}
{Albrow}, M.~D., {De Marchi}, G., \& {Sahu}, K.~C. 2002, \apj, 579, 660

\bibitem[{{Anders} \& {Fritze-v.~Alvensleben}(2003)}]{andersfritze03}
{Anders}, P. \& {Fritze-v.~Alvensleben}, U. 2003, \aap, 401, 1063

\bibitem[{{Baumgardt} \& {Makino}(2003)}]{baumgardt03}
{Baumgardt}, H. \& {Makino}, J. 2003, \mnras, 340, 227

\bibitem[{{Beasley} {et~al.}(2008){Beasley}, {Bridges}, {Peng}, {Harris},
  {Harris}, {Forbes}, \& {Mackie}}]{beasley08}
{Beasley}, M.~A., {Bridges}, T., {Peng}, E., {et~al.} 2008, \mnras, 386, 1443

\bibitem[{{Bekki} {et~al.}(2007){Bekki}, {Yahagi}, \& {Forbes}}]{bekki07}
{Bekki}, K., {Yahagi}, H., \& {Forbes}, D.~A. 2007, \mnras, 377, 215

\bibitem[{{Bertelli} {et~al.}(1994){Bertelli}, {Bressan}, {Chiosi}, {Fagotto},
  \& {Nasi}}]{bertelli94}
{Bertelli}, G., {Bressan}, A., {Chiosi}, C., {Fagotto}, F., \& {Nasi}, E. 1994,
  \aaps, 106, 275

\bibitem[{{Boutloukos} \& {Lamers}(2003)}]{boutloukos03}
{Boutloukos}, S.~G. \& {Lamers}, H.~J.~G.~L.~M. 2003, \mnras, 338, 717

\bibitem[{{Bruzual} \& {Charlot}(2003)}]{bruzual03}
{Bruzual}, G. \& {Charlot}, S. 2003, \mnras, 344, 1000

\bibitem[{{Dabringhausen} {et~al.}(2008){Dabringhausen}, {Hilker}, \&
  {Kroupa}}]{dabringhausen08}
{Dabringhausen}, J., {Hilker}, M., \& {Kroupa}, P. 2008, \mnras, 386, 864

\bibitem[{{Dubath} \& {Grillmair}(1997)}]{dubath97}
{Dubath}, P. \& {Grillmair}, C.~J. 1997, \aap, 321, 379

\bibitem[{{Gieles} {et~al.}(2005){Gieles}, {Bastian}, {Lamers}, \&
  {Mout}}]{gieles05a}
{Gieles}, M., {Bastian}, N., {Lamers}, H.~J.~G.~L.~M., \& {Mout}, J.~N. 2005,
  \aap, 441, 949

\bibitem[{{Girardi} {et~al.}(2000){Girardi}, {Bressan}, {Bertelli}, \&
  {Chiosi}}]{girardi00}
{Girardi}, L., {Bressan}, A., {Bertelli}, G., \& {Chiosi}, C. 2000, \aaps, 141,
  371

\bibitem[{{Harris}(1996)}]{harris96}
{Harris}, W.~E. 1996, \aj, 112, 1487

\bibitem[{{Harris} {et~al.}(2006){Harris}, {Whitmore}, {Karakla}, {Oko{\'n}},
  {Baum}, {Hanes}, \& {Kavelaars}}]{harris06}
{Harris}, W.~E., {Whitmore}, B.~C., {Karakla}, D., {et~al.} 2006, \apj, 636, 90

\bibitem[{{Hilker} {et~al.}(2007){Hilker}, {Baumgardt}, {Infante},
  {Drinkwater}, {Evstigneeva}, \& {Gregg}}]{hilker07}
{Hilker}, M., {Baumgardt}, H., {Infante}, L., {et~al.} 2007, \aap, 463, 119

\bibitem[{{Hillenbrand} \& {Hartmann}(1998)}]{hillenbrand98}
{Hillenbrand}, L.~A. \& {Hartmann}, L.~W. 1998, \apj, 492, 540

\bibitem[{{Kroupa}(2001)}]{kroupa01}
{Kroupa}, P. 2001, \mnras, 322, 231

\bibitem[{{Kruijssen} \& {Lamers}(2008)}]{kruijssen08}
{Kruijssen}, J.~M.~D. \& {Lamers}, H.~J.~G.~L.~M. 2008, \aap, submitted (KL08)

\bibitem[{{Lamers} {et~al.}(2006){Lamers}, {Anders}, \& {De Grijs}}]{lamers06}
{Lamers}, H.~J.~G.~L.~M., {Anders}, P., \& {De Grijs}, R. 2006, \aap, 452, 131

\bibitem[{{Lamers} {et~al.}(2005{\natexlab{a}}){Lamers}, {Gieles}, {Bastian},
  {Baumgardt}, {Kharchenko}, \& {Portegies Zwart}}]{lamers05}
{Lamers}, H.~J.~G.~L.~M., {Gieles}, M., {Bastian}, N., {et~al.}
  2005{\natexlab{a}}, \aap, 441, 117

\bibitem[{{Lamers} {et~al.}(2005{\natexlab{b}}){Lamers}, {Gieles}, \&
  {Portegies Zwart}}]{lamers05a}
{Lamers}, H.~J.~G.~L.~M., {Gieles}, M., \& {Portegies Zwart}, S.~F.
  2005{\natexlab{b}}, \aap, 429, 173

\bibitem[{{Mackey} \& {Gilmore}(2004)}]{mackey04}
{Mackey}, A.~D. \& {Gilmore}, G.~F. 2004, \mnras, 355, 504

\bibitem[{{Mandushev} {et~al.}(1991){Mandushev}, {Staneva}, \&
  {Spasova}}]{mandushev91}
{Mandushev}, G., {Staneva}, A., \& {Spasova}, N. 1991, \aap, 252, 94

\bibitem[{{McLaughlin}(2000)}]{mclaughlin00}
{McLaughlin}, D.~E. 2000, \apj, 539, 618

\bibitem[{{McLaughlin} \& {van der Marel}(2005)}]{mclaughlin05}
{McLaughlin}, D.~E. \& {van der Marel}, R.~P. 2005, \apjs, 161, 304

\bibitem[{{Mieske} {et~al.}(2008){Mieske}, {Hilker}, {Jordan}, {Infante},
  {Kissler-Patig}, {Rejkuba}, {Richtler}, {Cote}, {Baumgardt}, {West},
  {Ferrarese}, \& {Peng}}]{mieske08}
{Mieske}, S., {Hilker}, M., {Jordan}, A., {et~al.} 2008, ArXiv:0806.0374

\bibitem[{{Mieske} \& {Kroupa}(2008)}]{mieske08a}
{Mieske}, S. \& {Kroupa}, P. 2008, \apj, 677, 276

\bibitem[{{Mora} {et~al.}(2007){Mora}, {Larsen}, \& {Kissler-Patig}}]{mora07}
{Mora}, M.~D., {Larsen}, S.~S., \& {Kissler-Patig}, M. 2007, \aap, 464, 495

\bibitem[{{Rejkuba} {et~al.}(2007){Rejkuba}, {Dubath}, {Minniti}, \&
  {Meylan}}]{rejkuba07}
{Rejkuba}, M., {Dubath}, P., {Minniti}, D., \& {Meylan}, G. 2007, \aap, 469,
  147

\bibitem[{{Searle}(1971)}]{searle71}
{Searle}, L. 1971, \apj, 168, 327

\bibitem[{{Vandenberg} {et~al.}(1990){Vandenberg}, {Bolte}, \&
  {Stetson}}]{vandenberg90}
{Vandenberg}, D.~A., {Bolte}, M., \& {Stetson}, P.~B. 1990, \aj, 100, 445

\end{thebibliography}

\end{document}